\documentstyle[12pt,axodraw]{article}
\begin{document}
\baselineskip 0.6cm

\begin{titlepage}

\begin{flushright}
UCB-PTH-02/06 \\
LBNL-49604 \\
\end{flushright}

\vskip 1.0cm

\begin{center}
{\Large \bf Unification of Weak and Hypercharge Interactions \\
  at the TeV Scale}

\vskip 1.0cm

{\large
Lawrence J.~Hall and Yasunori Nomura
}

\vskip 0.5cm

{\it Department of Physics, University of California,
                Berkeley, CA 94720, USA}\\
{\it Theoretical Physics Group, Lawrence Berkeley National Laboratory,
                Berkeley, CA 94720, USA}

\vskip 1.0cm

\abstract{
A realistic $SU(3)_C \times SU(3)_W$ unified theory is constructed 
with a TeV sized extra dimension compactified on the orbifold
$S_1/Z_2$, leaving only the standard model gauge group 
$SU(3)_C \times SU(2)_L \times U(1)_Y$ unbroken in the low energy 4D
theory.  The Higgs doublets are zero modes of bulk 
$SU(3)_W$ triplets and serve to normalize the hypercharge generator, 
apparently giving a tree-level prediction for the weak mixing angle: 
$\sin^2 \theta = 1/4$.  The orbifold boundary conditions imply 
a restricted set of $SU(3)_W$ gauge transformations: at an orbifold 
fixed point only the transformations of $SU(2)_L \times U(1)_Y$ are 
operative. This allows quarks to be located at this fixed point, 
overcoming the longstanding problem of how to incorporate 
matter in a unified $SU(3)_W$ theory.  However, in general this local, 
explicit breaking of $SU(3)_W$ symmetry, necessary for including quarks 
into the theory, destroys the tree-level prediction for the weak 
mixing angle. This apparent contradiction is reconciled by making 
the volume of the extra dimension large, diluting the effects of 
the local $SU(3)_W$ violation. In the case that the electroweak 
theory is strongly coupled at the cutoff scale of the effective
theory, radiative corrections to the weak mixing angle can be reliably 
computed, and used to predict the scale of compactification: 
$1 - 2~{\rm TeV}$ without supersymmetry, and in the region of 
$3 - 6~{\rm TeV}$ for a supersymmetric theory.  The experimental signature 
of electroweak unification into $SU(3)_W$ is a set of ``weak partners'' 
of mass $1/2R$, which are all electrically charged and are expected 
to be accessible at LHC.  These include weak doublets of gauge particles 
of electric charge $(++,+)$, and a charged scalar.  When pair produced, 
they yield events containing multiple charged leptons, missing large 
transverse energy and possibly Higgs and electroweak gauge bosons.}

\end{center}
\end{titlepage}

\section{Introduction}
\label{sec:intro}

The quest for unification of the forces of nature has been a dominant 
theme of particle physics for the last 30 years. The weak force acts 
only at short distances, and must apparently have a very different 
underlying origin from the electromagnetic force. The triumph of the
standard electroweak theory is to provide a common picture for these
forces \cite{Glashow:tr}. However, while both the weak and 
electromagnetic forces are understood to arise from 
the interaction of spin 1 gauge particles,
the standard model does not unify these interactions. The weak and
electromagnetic forces originate from two separate gauge forces: the
weak force based on the group $SU(2)_L$ and the hypercharge force based
on the gauge group $U(1)_Y$.

The most striking success in the unification of the gauge forces of 
nature occurs in grand unified theories based on $SU(5)$ or $SO(10)$
symmetries \cite{Georgi:1974sy, SO10}. These groups contain 
both the electroweak group, $SU(2)_L \times U(1)_Y$, 
and the group, $SU(3)_C$, of the QCD gauge interaction of
the strong force, so that there is a single coupling constant for all
three interactions. Furthermore, the quarks and leptons of a single
generation are unified into one ($SO(10)$) or two ($SU(5)$)
representations of the unified gauge symmetry. Such embeddings of the
quarks and leptons force a unique normalization for hypercharge, so
that the relative strengths of the weak and
hypercharge forces is determined, leading to a tree-level prediction
for the weak mixing angle of $\sin^2\theta = 3/8$. Including
radiative corrections \cite{Georgi:1974yf}, a successful result 
follows only if weak scale supersymmetry is incorporated 
into the theory \cite{Dimopoulos:1981zb, Dimopoulos:1981yj},
in which case the scale at which the three forces are unified is found
to be of order $10^{16}$ GeV. This successful prediction has led to a
dominant paradigm for physics beyond the standard model: supersymmetry
at the TeV scale above which there is large energy desert, with no new 
physics appearing until $10^{16}$ GeV. Given this large energy, and
uncertainties in the nature of the grand unified theory, it has not been
possible to devise definitive experimental tests for this picture of force
unification. The new gauge bosons and scalars may simply be too heavy to 
give observable signals.

The first attempt to unify any of the gauge forces of nature came
before grand unification. It was an attempt at electroweak
unification, with the weak and hypercharge groups $SU(2)_L \times U(1)_Y$
unified into $SU(3)_W$ \cite{Weinberg:1971nd}. The theory 
possessed a hierarchy of symmetry breaking, 
with the unified symmetry breaking, $SU(3)_W \rightarrow 
SU(2)_L \times U(1)_Y$, occurring at a much larger mass scale
than the scale at which electroweak symmetry
breaks to electromagnetism $SU(2)_L \times U(1)_Y \rightarrow U(1)_{EM}$.
Embedding a lepton doublet $l=(\nu,e)_L$ and a lepton singlet
$e=\bar{e}_R$ into a fundamental representation of $SU(3)_W$, leads to the
tree-level prediction of $\sin^2\theta = 1/4$. While this is within 10\%
of the present experimental value of 0.231, this approach to gauge
unification was not pursued because it met immediate and insuperable
obstacles \cite{Weinberg:1971nd, Georgi:1972hy}. 
The most devastating difficulty is that
quarks cannot be accommodated in the theory. The quark doublet
$q=(u,d)_L$ has too small a hypercharge quantum number to appear in
any $SU(3)_W$ multiplet. Furthermore, the lepton assignment has gauge
anomalies, requiring the introduction of additional light charged 
fermions. One possibility is to build theories based on the
electroweak group $SU(3)_W \times U(1)$, but this clearly does not
unify the forces and does not predict the weak mixing angle. There
appears to be a fundamental inconsistency between electroweak
unification into $SU(3)$ and the observed quark and lepton quantum
numbers. 

In a recent proposal it has been demonstrated that the intriguing 
tree-level prediction of $\sin^2\theta = 1/4$ can be preserved 
even when the weak and hypercharge forces are not unified, provided
they are embedded in some larger semi-simple group that includes a new
$SU(3)'$ gauge interaction \cite{Dimopoulos:2002mv}. For example, 
at high energies assume the gauge forces of nature are based on
the expanded group $SU(4)_C \times SU(2)'_L \times SU(2)'_R \times
SU(3)'$, with quarks and leptons transforming in the usual way under
the Pati-Salam group, but not transforming under $SU(3)'$.
The Pati-Salam group is broken in the usual way to $SU(3)_C \times
SU(2)'_L \times U(1)'_Y$, and a further
symmetry breaking $SU(3)' \times SU(2)'_L \times U(1)'_Y
\rightarrow SU(2)_L \times U(1)_Y$ occurs in such a way that the
electroweak symmetry group, $SU(2)_L \times U(1)_Y$, is embedded partly
in $SU(3)'$ and partly in $SU(2)'_L \times U(1)'_Y$. 
The relation $\sin^2\theta = 1/4$ now emerges 
in the limit that the coupling constants for 
$SU(2)'_L$ and $SU(2)'_R$ are taken much larger than the
coupling for $SU(3)'$.

In this paper we do not follow this approach of adding a new $SU(3)'$
gauge interaction which does not couple to quarks and leptons. Rather, 
we return to the original idea of unifying the weak and hypercharge 
interactions into a single $SU(3)_W$ gauge force \cite{Weinberg:1971nd}, 
so that the standard model gauge group is embedded in 
$SU(3)_C \times SU(3)_W$, which becomes the symmetry group of
nature above the TeV scale. We use tools developed in 
Refs.~\cite{Hall:2001pg, Hall:2001xb}. 
Consider a higher dimensional theory with gauge group $G$ 
compactified on an orbifold, with different gauge fields having
different boundary conditions. This results in a theory with a
restricted set of gauge transformations; in particular, at orbifold
fixed points the operative gauge symmetry is $H$, a subgroup of
$G$. These points of reduced symmetry are very interesting. They can
support brane fields in any multiplets of $H$, whether these 
are parts of $G$ multiplets or not.  Similarly, the zero modes of bulk 
fields do not fill out complete $G$ multiplets.  At first sight, the 
presence of such points destroys the gauge coupling relation 
coming from $G$, since gauge kinetic terms that are not 
$G$ universal can be placed at the fixed points. 
However, this point defect breaking of gauge coupling unification is a 
small effect if the volume of the bulk is large compared to the cutoff 
of the effective higher dimensional theory. Said differently, the brane 
kinetic energy operators have higher mass dimension, and are therefore 
less relevant in the infrared, than the bulk kinetic energy operators. 
The larger the energy interval from the compactification scale up to 
the cutoff of the effective theory, the more accurate gauge coupling 
unification becomes. This powerful tool implies that, in higher 
dimensional field theories, it is possible to have
incomplete multiplets of the gauge group while still maintaining
tree-level gauge coupling unification. 
When applied to grand unification, it allows the construction of
completely realistic theories above the supersymmetric desert, which 
we call Kaluza-Klein (KK) grand unification. The minimal $SU(5)$ theory 
in 5D has automatic doublet-triplet splitting, no proton decay 
from dimension four or five operators, no unwanted mass relations 
for light generations, and a prediction for the QCD coupling of 
$\alpha_s = 0.118 \pm 0.005$ from gauge coupling 
unification \cite{Hall:2001xb}.

It is this new tool that allows us to overcome 
the obstacles to building an electroweak theory based
on $SU(3)_W$. The quarks and leptons do not need to fill out complete
$SU(3)_W$ multiplets if they live at orbifold fixed points. Because
the breaking of $SU(3)_W$ is localized, if the bulk has a large volume
the tree-level prediction, $\sin^2\theta = 1/4$, is maintained to
high accuracy. The normalization of the hypercharge generator within
the $SU(3)_W$ group is determined by the Higgs field which is a bulk
field transforming as a triplet of $SU(3)_W$. Including radiative
corrections to $\sin^2\theta$, the compactification scale can be
computed to the factor of three level, and is found to
be about 2 TeV, so that the new charged gauge bosons of $SU(3)_W$
and the new charged Higgs scalar are likely to be accessible to the LHC.

\section{The Basic Idea}
\label{sec:basic}

In this section we present the basic idea of our scheme.  We consider 
an $SU(3)_W$ gauge theory in 5D, compactified on an $S^1/Z_2$ orbifold.
Under the compactification, the gauge fields $A_M = \{ A_\mu, A_5 \}$ 
($\mu=0,\cdots,3$) obey the following boundary conditions:
\begin{eqnarray}
  A_\mu(x^\mu,y) = Z A_\mu(x^\mu,-y) Z^{-1} = 
    T A_\mu(x^\mu,y+2\pi R) T^{-1},
\label{eq:bc-Amu} \\
  A_5(x^\mu,y) = - Z A_5(x^\mu,-y) Z^{-1} = 
    T A_5(x^\mu,y+2\pi R) T^{-1},
\label{eq:bc-A5}
\end{eqnarray}
where $Z$ and $T$ are $3 \times 3$ matrices, and $A_M \equiv A_M^a T^a$ 
are also represented as $3 \times 3$ matrices.  To reduce the gauge group 
to $SU(2)_L \times U(1)_Y$ at low energies, we have two independent choices: 
(i) $\{ Z, T \} = \{ \mbox{diag}(1,1,1), \mbox{diag}(1,1,-1) \}$ and 
(ii) $\{ Z, T \} = \{ \mbox{diag}(1,1,-1), \mbox{diag}(1,1,1) \}$.
In the former case we have only $SU(2)_L \times U(1)_Y$ gauge fields, 
$A_\mu^{\rm EW}$, as massless fields, but in the latter case we also have 
off-diagonal pieces of the fifth component of the gauge fields, 
$A_5^X$.  Although it is possible to construct realistic theories 
based on the second choice, we defer the discussion of this possibility 
to section \ref{sec:alt} and here adopt the first one. Then, the KK 
tower for the gauge fields is given as follows: the standard-model 
$SU(2)_L \times U(1)_Y$ gauge bosons $A_\mu^{\rm EW}$ with masses $n/R$, 
joined at $n\neq 0$ levels by $A_5^{\rm EW}$, and the broken 
$SU(3)_W$ vectors $A_\mu^X$, joined by $A_5^{X}$, with masses 
$(n+1/2)/R$, where $n = 0,1,\cdots$.

\begin{figure}
\begin{center}
    \input{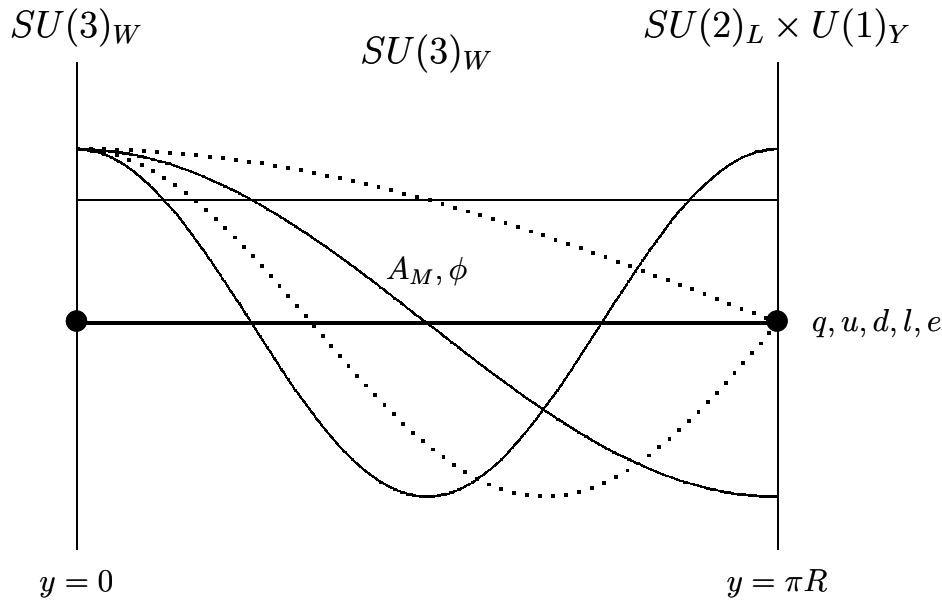}
\caption{The structure of the fifth dimension. Solid and dotted lines 
represent the profiles of gauge transformation parameters 
$\xi^{\rm EW}$ and $\xi^X$, respectively.  The gauge symmetry $SU(3)_W$ 
is reduced to $SU(2)_L \times U(1)_Y$ on the $y = \pi R$ brane, so that 
we can introduce quark and lepton fields on this brane.  The Higgs 
field $\phi$ is located in the bulk, fixing $\sin^2\theta = 1/4$ 
at tree level.}
\label{fig:orbifold}
\end{center}
\end{figure}

What is the symmetry of this system?  Since we are considering an effective 
field theory below the cutoff scale, it
makes sense to consider field theoretic symmetries using the classical 
spacetime picture.  We then find that the gauge symmetry of the system 
is $SU(3)_W$ but with the gauge transformation parameters obeying the 
same boundary conditions as the gauge fields:
\begin{equation}
  \xi(x^\mu,y) = Z \xi(x^\mu,-y) Z^{-1} = 
    T \xi(x^\mu,y+2\pi R) T^{-1},
\label{eq:bc-xi}
\end{equation}
which we refer to as restricted gauge symmetry \cite{Hall:2001pg}.
This means that while the $SU(2)_L \times U(1)_Y$ gauge parameters, 
$\xi^{\rm EW}$, have profiles $\cos[ny/R]$ in the extra dimension, 
$SU(3)_W/(SU(2)_L \times U(1)_Y)$ ones, $\xi^X$, have different 
profiles $\cos[(n+1/2)y/R]$, as depicted in Fig.~\ref{fig:orbifold}.
The crucial observation is that $\xi^X$ always vanish at $y=\pi R$, 
and hence the gauge symmetry is reduced to $SU(2)_L \times U(1)_Y$ 
on this point.  In particular, we can introduce any 
representation of $SU(2)_L \times U(1)_Y$ on the $y=\pi R$ brane, 
even if it does not arise from any $SU(3)_W$ representation.
This allows us to introduce quark and lepton fields, 
$q({\bf 2}, \alpha /6)$, $u({\bf 1}, -2 \alpha /3)$, 
$d({\bf 1}, \alpha /3)$, $l({\bf 2}, - \alpha /2)$, and 
$e({\bf 1}, \alpha)$, on this brane, where the numbers in parentheses 
represent $SU(2)_L \times U(1)_Y$ quantum numbers.  At this 
stage, the overall normalization of these $U(1)_Y$ charges, described 
by the parameter $\alpha$, is arbitrary, since it is not related to 
$SU(2)_L$ (or $SU(3)_W$) by the operation of $\xi^X$.

Does the explicit breaking of $SU(3)_W$ at $y= \pi R$
destroy the electroweak gauge coupling unification 
that originates from $SU(3)_W$ symmetry?  Generically, the answer is yes. 
We can write down gauge kinetic operators on the $y=\pi R$ brane with 
non-unified coefficients for $SU(2)_L$ and $U(1)_Y$ gauge fields.
However, since this $SU(3)_W$ breaking is point-like in the extra 
dimension, we can reduce its effect by making the volume of the extra 
dimension large, recovering the unified relation \cite{Hall:2001pg}.
Specifically, the most general form for the gauge 
kinetic energy is given by
\begin{equation}
  S = \int d^4x \; dy \; 
    \biggl[ \frac{1}{g_5^2} F^2 + 
    \delta(y - \pi R) \frac{1}{\tilde{g}_i^2} F_i^2 \biggr],
\label{eq:gaugekinops}
\end{equation}
where the first term is an $SU(3)_W$-invariant bulk gauge kinetic energy, 
while the second term represents non-unified kinetic operators 
located on the $y=\pi R$ brane ($i$ represents $SU(2)_L$ and 
$U(1)_Y$).  Here, we have omitted an $SU(3)_W$-invariant term on the 
$y=0$ brane, since it is irrelevant for the discussion below.
The zero-mode gauge couplings are obtained by integrating 
over the extra dimension:
\begin{equation}
  \frac{1}{g_i^2} = \frac{\pi R}{g_5^2} + \frac{1}{\tilde{g}_i^2},
\label{eq:4d-gc}
\end{equation}
where $g_1 = \sqrt{3}g'$ and $g_2 = g$.
Now, suppose the bulk and brane kinetic terms have ``comparable'' strength 
at the cutoff scale $M_s$.  This implies that $g_5 \approx \tilde{g}_i$ 
in units of $M_s$, so we write $\tilde{g}_i^2 = g_5^2 M_s/c_i$ 
and get
\begin{equation}
  \frac{1}{g_i^2} = \frac{\pi R}{g_5^2} 
    \left( 1 + \frac{c_i}{\pi R M_s} \right),
\end{equation}
where $c_i$ are non-universal coefficients of order unity.
We find that the $SU(3)_W$-violating effect from the $y=\pi R$ brane 
is suppressed by the volume of the extra dimension, $\pi R M_s$.
Therefore, by making the extra dimension large, we can reconcile two 
seemingly contradicting ideas: having fields that do not fit into 
any $SU(3)_W$ representation and keeping the $SU(3)_W$ relations for 
the gauge couplings!

Next we introduce the Higgs field as a bulk scalar field, $\phi$, 
transforming as a triplet under $SU(3)_W$.  We choose boundary 
conditions so that the $SU(2)_L$ doublet component of $\phi$ remains 
massless at tree level:
\begin{equation}
  \phi(x^\mu,y) = Z \phi(x^\mu,-y) = T \phi(x^\mu,y+2\pi R),
\label{eq:bc-phi}
\end{equation}
which gives a KK tower of masses $n/R$ for the doublet component, 
$\phi_D$, and a tower of masses $(n+1/2)/R$ for the singlet component, 
$\phi_S$.  We identify the zero mode of $\phi_D$ as the Higgs doublet, 
$h$, of the standard model.  This identification fixes the normalization 
of $U(1)_Y$.  Normalizing $U(1)_Y$ charges so that the Higgs doublet 
has $h({\bf 2}, -1/2)$, the $SU(2)_L$ and $U(1)_Y$ gauge couplings, 
$g$ and $g'$, are related by $g = \sqrt{3}g'$, giving $\sin^2\theta = 1/4$, 
at tree level.  In order that the usual Yukawa couplings, 
${\cal L}_4 = q u h^\dagger + q d h + l e h$, can be introduced on 
the $y=\pi R$ brane, the overall scale for the quark and lepton 
$U(1)_Y$ charges must be chosen to be $\alpha = 1$. 
Note that the Yukawa couplings respect only the 
$SU(2)_L \times U(1)_Y$ invariance and not the full $SU(3)_W$ symmetry.
The $SU(3)_C$ gauge fields are introduced either in the 5D bulk 
or on the $y=\pi R$ brane.  Below the compactification scale $M_c = 1/R$, 
the theory reduces to the standard model with $\sin^2\theta \approx 1/4$.
Radiative corrections to $\sin^2\theta$ are discussed in the 
next section.

\section{Calculable Framework}
\label{sec:calculable}

In the previous section, we have described a basic mechanism of 
introducing quarks and leptons while preserving the 
$SU(3)_W$ relation among the gauge couplings.  However, the correction 
to this relation from brane kinetic operators depends on the size 
of the extra dimension, which is an unknown free parameter.
Furthermore, although unlikely, it is a logical possibility 
that the coefficients of the brane operators 
are anomalously large, destroying the gauge coupling 
relation.  In Ref.~\cite{Hall:2001xb}, we have introduced a framework 
which removes these concerns and allows a reliable calculation 
for gauge coupling unification.  The crucial new ingredient is the 
assumption that the gauge interaction is strongly coupled at the 
cutoff scale $M_s$.  This assumption gives the largest possible volume 
for the extra dimension, and hence minimizes unknown contributions from 
tree-level brane operators.  It also determines the size of the leading 
radiative correction coming from the energy interval between $M_c$ 
and $M_s$.

To see how this works explicitly, let us first consider the effective 
action at the scale $M_s$.  No matter what physics occurs above $M_s$, 
the restricted gauge symmetry ensures that the 5D bulk is $SU(3)_W$ 
symmetric and all $SU(3)_W$-violating effects appear only on the 
$y=\pi R$ brane.  Therefore, the gauge kinetic energy must take 
the form of Eq.~(\ref{eq:gaugekinops}).  Now, since the theory 
is assumed to be strongly coupled at $M_s$, both bulk and brane 
gauge couplings are reliably estimated as $g_5^2 \simeq 16\pi^3/M_s$ 
and $\tilde{g}_i^2 \simeq 16\pi^2$ using naive dimensional 
analysis.\footnote{
One way of estimating these couplings is to consider loop diagrams 
in the equivalent 4D KK theory.  In the 4D picture, the bulk term gives 
gauge kinetic terms with KK momentum conservation, while the brane ones 
give terms with KK momentum violation.  After diagonalizing these kinetic 
terms, the gauge couplings among KK towers are obtained.
Requiring that contributions from all loop diagrams become comparable 
at the scale $M_s$ (i.e. the theory is strongly coupled at $M_s$), we 
obtain the result in the text, neglecting group theoretical 
factors of order unity.}
Inserting these estimates into Eq.~(\ref{eq:4d-gc}), we find that the bulk 
term contributes to $1/g_i^2$ an amount $M_s R/16\pi^2$ while 
the brane terms contribute an amount $1/16\pi^2$.  From this, we learn 
two things. First, since the 4D gauge couplings are $O(1)$, the volume 
of the extra dimension must be large, $M_s R = O(100)$.  Second, since 
the non-universal contribution is suppressed to the percent level, 
the resulting uncertainty in the calculation of gauge coupling 
unification is small, $\delta\sin^2\theta/\sin^2\theta \simeq 0.4\%$.

Having obtained gauge coupling unification at tree level at $M_s$, 
we turn to the quantum effects below $M_s$ that result 
from the $y=\pi R$ brane.  Below $M_s$, the 4D gauge couplings run 
by power law.  However, since the leading power-law piece comes from 
the evolution of the bulk term, it is $SU(3)_W$ symmetric.
Therefore, the relative running between $g$ and $g'$, which is 
relevant for gauge coupling unification, entirely 
comes from the evolutions of the gauge kinetic terms localized 
on the $y=\pi R$ brane.  Since these evolutions are logarithmic, they 
can be reliably computed in the effective theory.  Furthermore, they 
contribute to $1/g_i^2$ an amount $(1/16 \pi^2) \ln(M_s/M_c)$ 
and dominate the unknown tree-level correction of order $1/ 16 \pi^2$, 
by a factor of $\ln(M_s/M_c) \simeq \ln(100) \simeq 5$.
Including the radiative correction below $M_c$, we obtain the 4D 
gauge couplings at the weak scale:
\begin{equation}
  \frac{1}{g_i^2}(m_Z) \simeq \frac{1}{g_*^2} 
    + \frac{b}{4 \pi^2} \left[ \left(\frac{M_s}{M_c}\right) - 1 \right]
    + \frac{\tilde{b}_i}{8 \pi^2} \ln\frac{M_s}{M'_c}
    + \frac{b'_i}{8 \pi^2} \ln\frac{M'_c}{m_Z},
\label{eq:running}
\end{equation}
where $b$ and $\tilde{b}_i$ are the $\beta$-function coefficients 
above $M_c$, and $b'_i$ those below $M_c$; $g_*$ is the unified gauge 
coupling at $M_s$.  Here, we have matched the logarithmic 
contribution in higher dimensions to that in 4D at the scale 
$M'_c = M_c/\pi$, which represents the length scale of extra 
dimensions \cite{Nomura:2001mf}.  Eliminating $1/g_*^2$, we obtain 
the expression for $\sin^2\theta$ at $m_Z$:
\begin{equation}
  \sin^2\theta \simeq \frac{1}{4} - \frac{3}{8\pi} \alpha_{\rm em}
    \left[ (\tilde{b}_1-\tilde{b}_2) \ln\frac{M_s}{M'_c} 
    + (b'_1-b'_2) \ln\frac{M'_c}{m_Z} \right],
\label{eq:sin2theta}
\end{equation}
where $\alpha_{\rm em} \equiv e^2/4\pi \simeq 1/128$ represents the 
fine structure constant at $m_Z$.  Since the strong coupling requirement 
determines $M_s/M_c' \simeq 16\pi^3$, we can use this equation 
to estimate the compactification scale from the observed value of 
$\sin^2\theta$.  Assuming that the tree-level spectrum is not much 
changed by radiative corrections, we obtain the $\beta$-function 
coefficients above $M'_c$ as $(\tilde{b}_1, \tilde{b}_2) = (9/4, 1/4)$, 
following the prescription given in Ref.~\cite{Hall:2001xb}. 
(The gauge, Higgs and matter fields contribute $(0, -23/6)$, 
 $(1/36, 1/12)$ and $(20/9, 4)$, respectively).
Using the standard-model $\beta$-functions below $M'_c$, 
$(b'_1, b'_2) =  (41/18, -19/6)$, the compactification scale is 
estimated to be $1/R \simeq 1 - 2$ TeV.  It is interesting to note that 
both logarithmic runnings above and below $M_c$ reduce the value 
of $\sin^2\theta$ from $1/4$ with comparable contributions.  
Thus, even if we do not assume strong coupling at $M_s$, 
we expect the compactification scale to be around a TeV, as long as 
unknown contributions from tree-level brane operators are 
sufficiently small.

Since the TeV scale extra dimension suggests $M_s \approx 100$ TeV,
the present model needs a fine tuning to get 
$\left\langle h \right\rangle \equiv v \simeq 175$ GeV.  The required 
fine tuning is of order $v^2 / M_s^2 \approx 10^{-6}$.  However, this 
unpleasant feature is avoided by making the theory supersymmetric.
It is straightforward to supersymmetrize the model of the previous 
section.  The $SU(3)_W$ gauge field is now a 5D gauge supermultiplet, 
which consists of a 5D vector field, $A_M$, two gauginos, $\lambda$ and 
$\lambda'$, and a real scalar $\sigma$.  Using the 4D $N=1$ 
superfield language, $V(A_\mu, \lambda)$ and 
$\Sigma((\sigma+iA_5)/\sqrt{2},\lambda')$, the boundary conditions are 
given by Eqs.~(\ref{eq:bc-Amu}, \ref{eq:bc-A5}) with 
$A_\mu \rightarrow V$ and $A_5 \rightarrow \Sigma$.
The matter fields on the $y=\pi R$ brane become chiral 
superfields: $Q, U, D, L$ and $E$.  Since the Yukawa couplings 
on the brane must be supersymmetric, we need two Higgs multiplets.
Thus we introduce two Higgs hypermultiplets, $\{ \Phi, \Phi^c \}$ and 
$\{ \bar{\Phi}, \bar{\Phi}^c \}$ in the bulk, where $\Phi({\bf 3})$, 
$\Phi^c({\bf 3}^*)$, $\bar{\Phi}({\bf 3}^*)$, and $\bar{\Phi}^c({\bf 3})$ 
are 4D $N=1$ chiral superfields with $SU(3)_W$ transformations 
given in the parentheses.  For $\Phi$ and $\bar{\Phi}$, the boundary 
conditions are given by Eq.~(\ref{eq:bc-phi}) with 
$\phi \rightarrow \Phi, \bar{\Phi}$; for the conjugate fields, 
the boundary conditions are Eq.~(\ref{eq:bc-phi}) with 
$Z \rightarrow -Z$ and $\phi \rightarrow \Phi^c, \bar{\Phi}^c$.
These boundary conditions yield zero modes for $SU(2)_L$ doublets 
of $\Phi$ and $\bar{\Phi}$, which we identify with the two Higgs 
doublets of the minimal supersymmetric standard model (MSSM), 
$H_u \equiv \bar{\Phi}_{D,0}$ and $H_d \equiv \Phi_{D,0}$. The Yukawa 
couplings, $[Q U \bar{\Phi}_D + Q D \Phi_D + L E \Phi_D]_{\theta^2}$, 
are introduced on the $y=\pi R$ brane, and the QCD gauge interaction 
is introduced either in the 5D bulk or on the $y=\pi R$ brane.
Below $M_c$, the theory reduces to the MSSM with 
$\sin^2\theta \approx 1/4$.

As in the case of the non-supersymmetric theory, we can reliably 
estimate the compactification scale by requiring that the theory is 
strongly coupled at $M_s$.  Assuming that all superparticle masses 
are around $1/R$, we can use the standard-model $\beta$-functions 
below $M'_c$. The $\beta$-functions coefficients above $M'_c$ are 
given by $(\tilde{b}_1, \tilde{b}_2) = (10/3, 2)$.  (The gauge, 
Higgs and matter fields contribute $(0, -4)$, $(0, 0)$ 
and $(10/3, 6)$, respectively).  Then, using 
Eq.~(\ref{eq:sin2theta}), we obtain $1/R \approx 3$ TeV from 
the observed value of $\sin^2\theta$.  This estimate has a
considerable uncertainty coming from the actual superparticle spectrum; 
for example, in the extreme case that all superpartners are at $M_Z$, 
we find $1/R \approx 30$ TeV, using MSSM beta functions for $b_i'$.

While there are many possibilities for supersymmetry breaking, two
schemes are particularly well suited to our theory. The first
possibility is that of Scherk-Schwarz breaking of supersymmetry by
boundary conditions in the fifth dimension \cite{Antoniadis:1990ew}. 
In this case the gauginos and Higgsinos acquire tree level masses, 
while those of the quark and lepton superpartners arise at 
radiative level. The gauginos acquire mass $\approx 1/R$, while 
squarks and sleptons are much lighter.  Another natural possibility 
is that there is strong local breaking of supersymmetry on 
the $y=0$ brane \cite{Arkani-Hamed:2001mi}. This deforms the gaugino 
and Higgsino wavefunctions, giving them mass $1/R$, while the squarks 
and sleptons again acquire radiative masses.
In both these cases, with gauginos at $1/R$, the constraint from the
weak mixing angle leads to $1/R \approx 6$ TeV. 
Both schemes outlined here solve the supersymmetric flavor problem 
in an inherently extra-dimensional way. The locality of the squarks 
and sleptons forces supersymmetry breaking to be communicated to them 
via the gauge interactions.

Well beneath $M_c$, our theory reduces to the (supersymmetric) standard
model. The lightest states which signal the presence of $SU(3)_W$
electroweak unification are the lowest members of the $T$ odd KK
towers. These ``weak partners'' have mass $M_c/2 \approx 500 - 3000$ GeV, 
and are therefore expected to be within the reach of LHC. 
In the non-supersymmetric case, there are
two weak partners: the charged scalar $\phi_S$ (the weak partner of
the Higgs doublet) and an $SU(2)_L$ doublet of massive vectors 
$A_{\mu,5}^X$ (the weak partners of the electroweak 
gauge bosons). Radiative corrections will lift the
degeneracy of these states, and if $T$ is conserved the lightest weak
partner (LWP) will be stable. Since all the weak partners are
integrally charged under electromagnetism, this would be a striking
signal. The radiative corrections to the mass of $A_{\mu,5}^X$ are 
under control, because at short distances they are components of the 5D 
gauge field. However, the mass of the $\phi_S$, like that of the Higgs, 
is highly divergent, so its physical mass may be far from $M_c/2$.

While the bulk interactions necessarily preserve $T$, those on the
branes need not. The profiles of $\phi_S$ and $A_\mu^X$ vanish on the
$y= \pi R$ brane where the quarks and leptons are located, and hence
can only have derivative interactions with matter. The two fermion
interactions involve leptons but not quarks: $ll \partial_y \phi_S$
and $e^\dagger \sigma^\mu l \partial_y A_\mu^X$. While $A_5^X$ is
non-zero on this brane, its quantum numbers do not allow any couplings
to pairs of fermions. Thus, if $T$ violation is present, the LWP will
decay to leptons and not quarks. The LWP decay modes are 
$\phi^+ \rightarrow l^+ \nu$, $A^+ \rightarrow l^+ \nu$ and 
$A^{++} \rightarrow l^+ l^+$, where superscripts are electric charges, 
and $l^+$ is a charged lepton of any generation.
At LHC all accessible  weak partners will be pair produced via 
$s$-channel $\gamma, Z$ or $W^\pm$ exchange. The interaction 
$A_\mu^X \phi_S^\dagger \partial^\mu \phi_D$ allows cascade decays 
of the heavier of $\phi_S$ and $A_\mu^X$ to the lighter and $\phi_D$, 
so that the final state may also contain Higgs bosons, 
$W^\pm$ or $Z$. Similarly, the heavier of $A^+$ and $A^{++}$ 
can $\beta$ decay to the lighter via a $W^\pm$. The production
of singly charged LWPs yields events containing $l^+ l^-$,
possibly of differing generation, with large amounts of missing
transverse energy. Pair production of
doubly charged LWPs produces events with four isolated charged 
leptons. Pair production of non-LWP states leads to cascade 
decays followed by the LWP decay. Hence these events will have 
additional Higgs or weak bosons relative to the LWP events.

In the supersymmetric case the phenomenology of the bosonic weak
partners is not greatly changed, with similar striking events with
multi charged leptons. However, the weak partner states now fill out
multiplets of $N=1$ supersymmetry.  They are: a weak doublet vector 
multiplet $(A_\mu^X,\lambda^X)$, a weak doublet chiral multiplet 
$(A_5^X,\lambda'^X)$, and chiral multiplets $\Phi_S(\phi_S,\psi_S)$ 
and $\bar{\Phi}_S(\bar{\phi}_S,\bar{\psi}_S)$ and their conjugates. 
The radiative corrections to the masses of all these weak partners is 
under control. There is the possibility of a brane mass of the form 
$\Phi_S \bar{\Phi}_S$, but this is also expected to be of order $M_c$.

\section{Alternative Model}
\label{sec:alt}

In this section we discuss an alternative model in our scheme of 
electroweak unified theories.  In the previous sections, we have taken 
the boundary conditions in which $SU(3)_W$ is broken by the orbifold 
translation: $\{ Z, T \} = \{ \mbox{diag}(1,1,1), \mbox{diag}(1,1,-1) \}$.
However, we could alternatively choose the boundary conditions which 
breaks $SU(3)_W$ by the orbifold reflection: 
$\{ Z, T \} = \{ \mbox{diag}(1,1,-1), \mbox{diag}(1,1,1) \}$.
In this case, the zero mode sector contains the fifth component of the 
gauge fields that transforms as an adjoint of $SU(2)_L \times U(1)_Y$, 
$({\bf 2}, 0) + ({\bf 1}, 0)$, in addition to the 
$SU(2)_L \times U(1)_Y$ gauge fields and the electroweak Higgs doublet.
Since these extra fields are scalars in the 4D picture, they 
receive masses of order $(1/4\pi)(1/R)$ through radiative corrections, 
and the model can be phenomenologically viable.  
An interesting property of this setup is that the gauge symmetry 
structure is different from that in the previous models.  
Specifically, $SU(2)_L \times U(1)_Y$ and 
$SU(3)_W/(SU(2)_L \times U(1)_Y)$ gauge parameters, 
$\xi^{\rm EW}$ and $\xi^X$, have profiles $\cos[ny/R]$ and $\sin[ny/R]$, 
respectively, so that the gauge symmetry is reduced to 
$SU(2)_L \times U(1)_Y$ on both $y=0$ and $y=\pi R$ branes. The Higgs field 
is located in the bulk as a triplet of $SU(3)_W$, so that it determines 
$\sin^2\theta = 1/4$ upon identifying the doublet component with $h$.
Then, we can put quarks, $q,u,d$, and leptons, $l,e$, on 
different branes; for example, quarks on $y=0$ and leptons on  
$y=\pi R$.  This is interesting because it provides 
proton stability through the separation of quarks and leptons in the 
extra dimension \cite{Arkani-Hamed:1999dc}.

Much of the collider phenomenology of this
theory is similar to that discussed above for our first theory. The
weak partners again consist of a charged scalar, $\phi_S$, and a heavy
vector doublet, $A^X_{\mu,5}$, but now have a mass $1/R$. 
While all the states of this theory are $T$ even, these weak partners
are $Z$ odd and potentially stable.
There is, however, a crucial new ingredient: there is a doublet of
scalars which are even under both $Z$ and $T$ and acquire a radiative
mass of order $(1/4\pi)(1/R)$. These states could be accessible to
both the Fermilab collider and the LHC. They will be distinctive since
they come with both $++$ and $+$ charges.  If the brane 
operator $ll \partial_y \phi_S$ is present, these $A^X_5$
scalars will decay via a virtual $\phi_S$ to $ll \phi_D$. Thus the
signal collider events contain either 2 or 4 isolated charged leptons, 
together with two electroweak gauge or Higgs bosons. 

\section{Conclusions}
\label{sec:concl}

We have proposed a unification of weak and hypercharge gauge forces at
the TeV scale into a single $SU(3)_W$ interaction. One motivation for
such a unification is a tree-level prediction of the weak mixing angle,
$\sin^2 \theta = 1/4$, but apparently there is a fatal difficulty: quarks
do not fit into $SU(3)_W$ multiplets. A new opportunity arises if
$SU(3)_W$ is realized as a gauge symmetry in 5D rather than in 4D.
The additional dimension is compactified on $S_1/Z_2$, with boundary
conditions inducing $SU(3)_W \rightarrow SU(2)_L \times U(1)_Y$. This
leads to a fixed point which does not respect the full $SU(3)_W$
symmetry, but only its $SU(2)_L \times U(1)_Y$ subgroup, allowing
quarks, which do not have the right quantum numbers to appear in
$SU(3)_W$ multiplets, to be located on this fixed point. However,
since $SU(3)_W$ is explicitly broken at this fixed point, it is not 
clear that the unification of hypercharge and weak gauge couplings
persists. We have argued that the local operators for gauge kinetic
terms which violate $SU(3)_W$ are higher dimensional and irrelevant, so
that the tree level prediction is preserved only in the case that the
bulk has a large volume. We have pursued the possibility that the
electroweak gauge sector is strongly coupled at high energies, and used
this assumption to predict the compactification scale: $1/R \approx 
1 - 2$ TeV without supersymmetry. The scale $1/R$ is much more
uncertain in the supersymmetric case, due to the superpartner
spectrum. In schemes for supersymmetry breaking which solve the
supersymmetric flavor problem, we find $1/R \approx 3 - 6$ TeV.
Experimental signatures for $SU(3)_W$ unification are provided by 
``weak partners'' with mass $1/2R \approx 500 - 3000$ GeV: 
a charged scalar partner of the Higgs doublet, and a weak doublet of
heavy gauge bosons. These states will be pair produced at LHC. If the
orbifold translation symmetry is unbroken the lightest weak partner
(LWP), which is charged, will be stable. If the translation symmetry
is broken, the decays of the weak partners lead to characteristic
events with several isolated charged leptons.

We have argued that there is a large energy interval above the TeV 
scale where the effective theory of nature is 5D with gauge 
group $SU(3)_C \times SU(3)_W$. If color propagates in 
the bulk, then a further unification of forces is possible at 
higher energies \cite{Dienes:1998vh}.  Both color and weak 
gauge couplings undergo power-law running with beta 
function coefficients given by $-3$ and $-2$ respectively. 
Since QCD is more asymptotically free, the couplings approach 
each other at high energies, and meet at $M_* \approx 4 \pi^2
M_c (1/g_W^2(M_c) - 1/g_C^2(M_c)) \approx 100$ TeV. Although this
calculation has very large uncertainties, coming from power-law
sensitivity to unknown ultraviolet physics, it is encouraging that
this suggested grand unification scale is close to the scale at which
the 5D theory becomes strongly coupled, $M_s$. At this higher scale of
order 100 TeV our effective 5D theory may be embedded into a more
fundamental theory, which includes gravity.\footnote{
One possible grand unified theory is $SU(6)$ in 6D, compactified on 
$T_2/(Z_2 \times Z_2')$, with $R_5 \gg R_6$. Boundary conditions in the 
$x_5$ direction break $SU(6) \rightarrow SU(5) \times U(1)$, while those 
in the $x_6$ direction break $SU(6) \rightarrow SU(3)_C \times SU(3)_W 
\times U(1)$. Below the grand unification scale of $1/R_6 \approx M_* 
\approx 100$ TeV, the effective 5D theory can be any of the models 
described in this paper (with an additional $U(1)$).}
For gravity to be strong at this scale requires some additional 
very large dimensions in which only gravity propagates 
\cite{Arkani-Hamed:1998rs}. Although the 100 TeV scale is somewhat 
higher than originally proposed, it comfortably avoids flavor problems, 
as well as cosmological and astrophysical limits on the case of 
two very large extra dimensions.

After the completion of this paper, we received Ref.~\cite{Li:2002} 
which considers $SU(3)_W$ in 5D.

\vspace{0.3cm}

{\bf Note added:}

In the models discussed above, the normalization of the quark and
lepton hypercharges relative to those of the Higgs, $\alpha$, is not
determined by the theory; rather $\alpha = 1$ was imposed as 
a phenomenological requirement. Here we construct theories in which 
this charge quantization ($\alpha = 1$) is determined by the 
consistency of the theory.\footnote{
We assume identical gauge quantum numbers for each generation.}
Since the leptons have the correct quantum numbers to fit into 
an $SU(3)_W$ triplet, they can appear on the $y=0$ brane, or in the bulk
as two triplets per generation having opposite $T$ quantum numbers 
\cite{Dimopoulos:2002}. In both cases, triplet leptons induce a brane
localized $SU(3)_W$ gauge anomaly. We find that this gauge anomaly at
$y=0$ can be cancelled by adding a bulk Chern-Simons term with 
fixed normalization, having a coefficient which is constant in
the bulk, $Z$ odd and $T$ even. This Chern-Simons term induces an
anomaly of fixed size on the $y= \pi R$ brane. The $SU(3)_W /
(SU(2)_L \times U(1)_Y)$ gauge fields have profiles which vanish at 
$y = \pi R$, so this anomaly is only for the $SU(2)_L \times U(1)_Y$ 
gauge fields. It is cancelled by a unique choice for the normalization 
of the hypercharges of the quarks. Thus gauge invariance is recovered for
the entire $SU(3)_W$ theory, and the overall normalization for both
quark and lepton hypercharges is determined  ($\alpha = 1$).
In these theories, our calculations of $1/R$ are unchanged. The
leptons fill complete $SU(3)_W$ multiplets, and do not affect relative
running. 

The phenomenology of the theory with bulk leptons is similar to that
discussed in section \ref{sec:calculable}. There are additional $T$ odd 
weak partner states of mass $1/2R$: $SU(2)_L$ doublet and singlet vector 
leptons $L$ and $E$. Brane interactions such as $Le \phi_D$ and 
$lE \phi_D$, which violate $T$, could also contribute to LWP decays. 
These operators could lead to single production of $L$ and $E$ in 
$e^+ e^-$ collisions. Since the leptons are bulk modes, their masses, 
arising from brane Yukawa couplings, are volume suppressed by $1/(M_s R)$
relative to quark masses; a trend observed in all three generations.
In the supersymmetric case the sleptons receive a tree level mass and
are heavier than the squarks, which only acquire a radiative mass.

The phenomenology of triplet leptons on the $y=0$ brane is altered
because the weak partner vector boson $A^X_\mu$ has $T$ violating 
interactions with the lepton current $e^\dagger l$, and the Higgs doublet 
must originate from an $SU(3)_W$ sextet \cite{Dimopoulos:2002}. Hence the
scalar weak partners are a weak triplet $\phi_T(--,-,0)$ and a weak
singlet $\phi'_S(++)$ which have $T$ violating couplings on the 
$y = 0$ brane of $ll \phi_T^\dagger + ee \phi'^\dagger_S$. The decays 
of the weak partners, as always, are to leptons and not to quarks. 
If light enough, the vector weak partners could be produced 
singly at future lepton colliders via 
$e_L^- e_R^- \rightarrow A_X^{--}, A_X^- W^-$, but not via $e^+ e^-$ 
annihilation.

\section*{Acknowledgements}

Y.N. thanks the Miller Institute for Basic Research in Science 
for financial support.  This work was supported in part by the Director, 
Office of Science, Office of High Energy and Nuclear Physics, of the U.S. 
Department of Energy under Contract DE-AC03-76SF00098, and in part 
by the National Science Foundation under grant PHY-00-98840.

\end{document}